\documentstyle[12pt]{article}
\newcommand\be{\begin{equation}}
\newcommand\ee{\end{equation}}
\newcommand\bea{\begin{eqnarray}}
\newcommand\eea{\end{eqnarray}}
\newcommand\ket[1]{|#1\rangle}

\newcommand\braket[2]{\langle #1|#2\rangle}
\newcommand{\fatalpha}{{\bf \alpha \kern -0.44em \alpha}}
\newcommand{\fatsigma}{{\bf \sigma \kern -0.54em \sigma}}
\newcommand{\tpchi}{{\bf \chi \kern -0.35em \chi}}
\newcommand{\llambda}{{\bf \lambda \kern -0.45em \lambda}}

\bibliography{plain}
\title{\bf Perfect state transfer via quantum probability theory }\vspace{20mm}
\author{ S. Salimi$^a$\thanks{Corresponding author:shsalimi@uok.ac.ir},
          S. Ghoraishipour$^a$\thanks{ghoraishipour@uok.ac.ir}
          and A. Sorouri$^a$\thanks{a.sorouri@uok.ac.ir}\\
          {$^a$\footnotesize \emph{Department of Physics, University of Kurdistan,}}\\
           {\footnotesize \emph{P.O.Box 66177-15175 , Sanandaj, Iran}.}\\
       }

\pagebreak

\vspace{20mm}
\begin{document}
\maketitle \vspace{15mm}

\begin{abstract}
The transfer of  quantum states has played an  important role in
quantum information processing. In fact, transfer of quantum states
from point $A$ to $B$ with unit fidelity is very important for us
and we focus on this case. In recent years, in represented works,
they designed Hamiltonian in a way that a mirror symmetry creates
with with respect to network center. In this paper, we stratify the
spin network with respect to an arbitrary vertex of the  spin
network ¤o¤ then we design coupling coefficient in a way to create a
mirror symmetry in Hamiltonian  with respect to center. By using
this Hamiltonian and represented approach, initial state that have
been encoded on  the first vertex in suitable time and with  unit
fidelity from it's antipodes vertex can be received. In his work,
there is no need to external control.


{\bf PACs Index: 03.65.Ud }
\end{abstract}

\vspace{70mm}
\newpage
\section{Introduction}
Transfer of a quantum state from one qubit to another is very
important for quantum information processing \cite{ekert}. In the
 quantum computers high quality  communication between different
parts of the system is essential, therefore we need to transfer
quantum states within the quantum processor. Depending on the
technology at hand, this task can be accomplished in a number of
ways. One of them is a quantum spin network which can be defined as
a collection of interacting quantum two-state systems ( or qubits )
on a graph. Spin network dynamics  is governed by a suitable
Hamiltonian such as  the Heisenberg or $XY$ Hamiltonian. Spin
networks do not require external controls during the transport of
information. It is an advantage  because any external control can be
a  source of noise.  A Perfect state transfer between two qubits of
a spin network is accomplished if a single excitation can travel
from one qubit to another with the fidelity being equal to 1.  Bose
is the  first one who suggested to use a 1D chain of $N$ qubits
permanently coupled which is described  by the nearest-neighbor
Heisenberg Hamiltonian \cite{Bose}
\begin{equation} H_{ij}=J_{ij}\vec{S_{i}}\cdot\vec{S_{j}}
\end{equation}
in
\begin{equation} H_{Bose}=J\sum_{n=1}^{N-1}( \sigma_{n}^{x}\sigma_{n+1}^{x} + \sigma_{n}^{y}\sigma_{n+1}^{y}
+\sigma_{n}^{z}\sigma_{n+1}^{z} ) \end{equation}
 where $\sigma_{n}^{i}, \sigma_{n+1}^{i}$ are the pauli matrices
 acting on the $n$-th and $(n+1)$-th qubit with $i=x,y,z,$
 and $J$ is the  coupling strength between qubits. The aim of the suggestion is to transfer a state from end of the
 chain to the other end after a definite time $\tau$.  Bose's idea has
motivated many studies focusing on the perfect quantum state
 transfer in spin networks [3-32]. For a spin chain with two and three qubits, transfer can be
achieved perfectly,  but for larger $N$, i.e. $N\geq 4$, it can be
shown that perfect transfer is impossible  \cite{cdel}. By
modulating the coupling and mirror symmetry of a spin chain,
Christandl \emph{et al}, have  shown that perfect state transfer can
be achieved for  spin chain with any $N$ . They  have also
demonstrated that how  spin networks can be used  to transfer
entangled quantum states and to generate entanglement between
different sites in the network. In Ref. \cite{chen}, Christandl's
studies on quantum states have been extended to the high-excitation
states. It is also shown that the entangled states in the form of $
\alpha \ket{00\cdots0}+ \beta\ket{11\cdots1} $  can be transferred
perfectly through a spin chain.

Star networks is an example of  spin networks which is potentially
useful for connecting different parts of a quantum network.
Investigation of the quantum state transfer in a star network and
generalization of the spin chain engineering problem to the topology
of a star network have been considered \cite{Yung 1}. For a 3-spin
chain the perfect state transfer has been experimentally tested by
using liquid nuclear magnetic resonance \cite{zhang}. It has been
suggested that the "dual-rail" encoding which adopts two parallel
quantum channels, can achieve arbitrary perfect quantum state
transfer \cite{bb1}. Authors in Ref.\cite{salimi} have introduced a
new method for calculating the probability amplitudes of quantum
walk based on spectral distribution. In this method a canonical
relation between the Fock space  of stratification graph and set of
orthogonal polynomials has been established which leads to obtain
the probability measure (spectral distribution) of adjacency matrix
graph. The method of spectral distribution only requires simple
structural data of graph and allows us to avoid a heavy
combinational argument often necessary to obtain full description of
spectrum of the adjacency matrix. In this paper,  we will consider spin networks as graphs in which the
vertices are qubits. Then, by using the method which has been introduced in the Ref .\cite{salimi}  we
design the spin network and can transfer a quantum state perfectly between antipodes vertices.

 The paper is organised as follows: In Sec. $2$,
we  provide a brief  review on the concept of the graph, adjacency
matrix, transfer of quantum states and fidelity. In Sec. 3, we look
at  the works in Ref.\cite{salimi} and its references. We also
introduce the concept of  the stratification and quantum
decomposition by which we can transfer quantum states between
various networks antipodes perfectly. To support our method, we
provide some interesting examples in Sec. 4. Finally, section 5 is
the conclusion.

\section{Graph, perfect state transfer between antipodes of spin
networks}
 A graph $G=(V,E)$  is generally defined  by two sets. One set ($V(G)$) is
  called the vertices set, including $N$ integer numbers
 from 1 to $N$, $V(G)=
\{1,2,\cdots N\}$. Another set is actually a subset ($E(G)$) of the
Cartesian product of the vertices set by itself, $E \subset V\times
V$. Two vertices $i,j \in V(G)$ are adjacent if and only if $(i,j)
\in E(G)$, and in this case we write  $i\sim j$. The degree  or
valency of a vertex $i\in V $ is defined by
\begin{equation}
 \kappa(i)=| \{j\in V | i\sim j\}|,
 \end{equation}
 where $|A|$ is the cardinality of the set $A$. Any graph can be characterized by adjacency matrix $A(G)$,
\[
A_{i j}(G) = \left\{
\begin{array}{ll}
1 & \mbox{if $ (i,j)\in E(G)$}\\
0 & \mbox{otherwise.}
\end{array}
\right.
\]
Let us consider a system of $N$ spin-$1/2$ particles. Each particle
is represented by a vertex of $G$ and the Hilbert space associated
with $G$ is $\left(C^{2}\right)^{\otimes N }$, where $C$ is the set
of the complex numbers. We consider the Hamiltonian of a modified
Heisenberg XX model
\begin{equation} H_{G}=\sum_{(i,j)\in
E(G)}J_{i,j}(\sigma_{i}^{x}\sigma_{j}^{x}
+\sigma_{i}^{y}\sigma_{j}^{y}) ,\end{equation} Where $J_{i,j}$ is
coupling strength between vertices $i$ and $j$, and $\sigma_{i}^{l}$
and $\sigma_{j}^{l}$ ($l=x, y,z$) are the Pauli matrices acting on
the $i$-th and $j$-th vertices, respectively. The standard basis for
the 1-qubit Hilbert space can be the set $\{\ket{0}\equiv
\ket{\downarrow} \ , \ \ket{1}\equiv \ket{\uparrow}\}$. According to
the quantum state transfer protocol, a quantum state is written as
\begin{equation}
\ket{\psi_{in}}=\alpha \ket{\underline{0}}+\beta
\ket{\underline{1}},\;\;\;\;\;\;\
 |\alpha|^{2}+|\beta|^{2}=1,
 \end{equation}
where $\ket{\underline{0}}=\ket{0_{1}0_{2}\cdots0_{N}}$ represent a
network state in which all spin are down, and
$\ket{\underline{1}}=\ket{1_{1}0_{2}\cdots 0_{N}}$ shows a state
with the first spin up and the rest down. It is clear that the state
$\ket{\underline{0}}$ is an eigenvector of the Hamiltonian
corresponding to  zero-value of eigenvalue,
\begin{equation}
H_{G}\ket{\underline{0}}= E_{0}\ket{\underline{0}}, \;\;\;\;\;\;\
E_{0}=0 .\end{equation}
Thus, $\alpha$ does not change in time. Since the
total $\textit{z}$ -component of the spin by \begin{equation}
\sigma_{tot}^{z} = \sum_{i\in V(G)}\sigma_{i},\end{equation}
commutes with $H_{G}$, i.e. $[\sigma_{tot}^{z} , H_{G}]=0$.  From
this relation  can be
understood that initial state must evolve into single-excitation
space, in states that one spin is $\ket{\uparrow}$ state and all
other spins are $\ket{\downarrow}$ state. Define $U(t)=
e^{-iH_{G}t}$ as quantum mechanical time evolution operator, thus
network state in time $t$
\begin{equation}
\ket{\psi(t)}=\alpha\ket{\underline{0}}+\beta\sum_{n=1}^{N}
f_{n,1}(t)\ket{n},
\end{equation} where, $f_{n,1}(t) = \langle n |U_{G}(t)|1\rangle $
is the fidelity at time $t$ between $n$ and $1$. We say that there
is perfect state transfer between vertices $N$ and $1$ at time $t$
if and only if
\begin{equation}|f_{N,1}(t)|=1.
\end{equation}
To obtain this condition we use the method of the quantum probability theory which
 we discuss it in the next section.

\section{Quantum decomposition and spectral distribution of adjacency matrix}
 We define a walk of
length $k$ (or $k$ steps) for a finite sequence
$i_{0};i_{1};\cdots ; i_{k} \in V(G)$ if $i_{n-1}$ adjacent to
$i_{n}$ $(i_{n-1}\sim i_{n})$.
 Also in the Hilbert space, adjacent matrix acts as
following
\begin{equation} A\ket{i}=\sum_{i\sim j}\ket{j},\end{equation} and
$i\in V(G)$. Let us define $\partial(i,j)$ be the length of
shortest walk connecting $i$ and $j$ vertices. We note that
$\partial(i,i)= 0$ for all vertices. We  consider one vertex such
$o \in  V(G)$ as reference vertex and  stratify the spin network
(graph) with respect to $o$ vertex. Then, the graph can be
stratified into a disjoint union of strata
\begin{equation}
V=\bigcup_{k=0}^{\infty}V_{k}, \;\;\;\;\;\;\ V_{k}=\{i\in
V;\partial(o,i)=k\}.
\end{equation}
With each associate class
$V_{k}$ we associate a unit vector in Hilbert space defined
by\begin{equation}
\ket{\phi_{k}}=\frac{1}{\sqrt{|V_{k}|}}\sum_{i\in V_{k}}
\ket{i},\end{equation}where $\ket{i}$ denote the eigenket of the
$i$-th vertex at the stratum $k$.
 We define $\Gamma(G)$ as a closed subspace of Hilbert space
 spanned by $\{|\phi_{k}\rangle\}$, thus we can write
\begin{equation}\Gamma(G)=\sum_{k}\oplus\textbf{
 C}\ket{\phi_{k}}.\end{equation}
 Now we want to review the quantum decomposition of matrix adjacent associated with the stratification(3-12). We consider $A$ as a adjacency matrix of a graph $G=(V,E)$, then
using stratification that was explained above, we can define three
matrices $A^{-},A^{+} $ and $A^{0}$ as a follows and for $i\in
V_{k}$\begin{equation}A^{-}\ket{i}=\sum_{j\in
V_{k-1}}\ket{j},\;\;\;\ A^{0}\ket{i}=\sum_{j\in
V_{k}}\ket{j},\;\;\;\ A^{+}\ket{i}=\sum_{j\in
V_{k+1}}\ket{j},\end{equation} where $j\sim i$.  Since $i \in V_{k}$
and $i$ adjacent to $j$ then $j \in V_{k-1}\bigcup V_{k+1}\bigcup
V_{k}$. Thus we can
write\begin{equation}A=A^{-}+A^{0}+A^{+},\end{equation} (for more
details see Ref.\cite{obata}), we fix a vertex $o \in V$ as an
origin of the graph, called reference vertex and consider
$\ket{\phi_{o}}$
as a vector state $\ket{o}$, i.e. $\ket{o}=\ket{\phi_{o}}$.\\
According to Ref.\cite{obata}, $\langle A^{m} \rangle $ the
expectation value of power adjacency matrix $A$ with respect to a
state $\ket{\phi_{0}}$, coincides with the number of $m$-step walk
starting and terminating at $o$, then by using \cite{obata}, we
define two Szeg\"{o}-Jacobi sequences
$\left\{\omega_{k}\right\}_{k=1}^{\infty}$ and
$\left\{\alpha_{k}\right\}_{k=1}^{\infty}$, such
that\begin{equation}
A^{-}\ket{\phi_{k}}=\sqrt{\omega_{k}}\ket{\phi_{k-1}}, \;\;\;\
A^{-}\ket{\phi_{0}}=0, \;\;\;\ k\geq 1,\end{equation}
\begin{equation}A^{0}\ket{\phi_{k}}=\alpha_{k+1}\ket{\phi_{k}},
\;\;\;\ k \geq 0,\end{equation}
\begin{equation}
A^{+}\ket\phi_{k}=\sqrt{\omega_{k+1}}\ket{\phi_{k+1}}, \;\;\;\
k\geq 0,\end{equation} where $\sqrt{\omega_{k+1}}=
\frac{|V_{k+1}|^{1/2}}{|V_{k}|^{1/2}}\kappa_{-(j)},\kappa_{-(j)}=|\{i\in
V_{k}; i\sim j\}|$ for $j \in V_{k+1}$ and also $\alpha_{k+1}=
\kappa_{0(j)}, $such that $\kappa_{0(j)}=|\{i\in V_{k};i\sim j\}|$
for $j \in V_{k}$. Or, equivalently\begin{equation} A\ket
{\phi_{k}}=\sqrt{\omega_{k}}\ket{\phi_{k-1}}
+\alpha_{k+1}\ket{\phi_{k}}
+\sqrt{\omega_{k+1}}\ket{\phi_{k+1}},\end{equation} we note that
the $\ket{\phi_{-1}}=0$.

Now, by using the method of quantum decomposition we define the
spectral distribution of adjacency matrix \cite{salimi, obata}. The
spectral analysis of operators is an important issue in quantum
mechanics, operator theory and mathematical physics. It is
well-known, for any pair $(A,  \ket{\phi_0})$  of a matrix $A$ and a
vector $ \ket{\phi_0}$, one can be assigned a measure $\mu$ as
\begin{equation}\label{1}
\mu(x)=\langle \phi_0|E(x)|\phi_0\rangle,
\end{equation}
where  $E(x)=\sum_l |u_l\rangle \langle u_l|$ is the operator of
projection onto the eigenspace of $A$ corresponding to eigenvalue
$x$, i.e,
\begin{equation}\label{2}
A=\int x E(x)dx.
\end{equation}
Also, it is easy to see that for any polynomial $P(A)$ we have
\begin{equation}
P(A)=\int P(x) E(x)dx,
\end{equation}
where for discrete spectrum the above integrals are replaced with
summation.

Here we are interested to the spectral distribution of the adjacency
matrix of graphs, since the spectrum of a given spin network can be
determined by spectral distribution of its adjacency matrix $A$.

Therefore, by using relations (\ref{1}) and (\ref{2}), the
expectation value of the power  of adjacency matrix $A$  can be
written
\begin{equation}\label{3}
\langle A^m\rangle=\int_{R}x^{m}\mu(dx), \;\;\;\;\ m=0,1,2,...
\end{equation}
where $\langle . \rangle$ is the expectation value with respect to a
reference vector $\ket{\phi_{0}}$. The existence of the spectral
distribution which satisfy  Eq. (\ref{3}) is a consequence of
Hamburgers theorem \cite{Hamburgers}( see theorem 1.2).
 Therefore, the spectral distribution $ \mu $
under question will be characterized  by the property of
orthogonalizing polynomials $\{P_{n}\}$ defined recurrently by
$$ P_0(x)=1, \;\;\;\;\;\
P_1(x)=x-\alpha_1,$$
\begin{equation}\label{6}
xP_{n}(x)=P_{n+1}(x)+\alpha_{n+1}P_{n}(x)+\omega_nP_{n-1}(x),
\;\;\;\;\ n\geq1.\end{equation} If such a spectral distribution is
unique, the spectral distribution $\mu $ is determined by the
identity
\begin{equation}\label{4}
G_{\mu}(x)=\int_{R}\frac{\mu(dy)}{x-y}=\frac{1}{x-\alpha_{1}-\frac{\omega_{1}}{x-\alpha_{2}-\frac{\omega_{2}}
{x-\alpha_{3}-\frac{\omega_{3}}{x-\alpha_{4}-\cdots}}}}=\frac{P_{n-1}^{(1)}(x)}{P_{n}(x)}=\sum_{l=1}^{n}
\frac{A_l}{x-x_l},
\end{equation}
where, $x_{l}$ are the roots of polynomial $P_{n}$. $G_{\mu}(x)$
is called the Stieltjes transform, $A_{l}$ is the coefficient in
the Gauss quadrature formula and where the polynomials
$\{P_{n}^{1}\}$ are defined recurrently as
$$P_{0}^{(1)}(x)=1,\;\;\;\;\
    P_{1}^{(1)}(x)=x-\alpha_2, $$
   \begin{equation} xP_{n}^{(1)}(x)=P_{n+1}^{(1)}(x)+\alpha_{n+2}P_{n}^{(1)}(x)+\omega_{n+1}P_{n-1}^{(1)}(x), \;\;\;\;\ n\geq1. \end{equation}
Then, the spectral distribution $\mu$ can be recovered from
$G_{\mu}(x)$ by means of the Stieltjes inversion formula:
    \begin{equation}\label{5}
\mu(y)-\mu(x)=-\frac{1}{\pi}\lim_{v\longrightarrow
0^+}\int_{x}^{y}Im\{G_{\mu}(u+iv)\}du.
\end{equation}
Substituting the right hand side of Eq.(\ref{4}) in the Eq.(\ref{5}), the spectral
distribution $\mu$ can be determined in terms of
$x_{l},l=1,2,\cdots $, and Guass quadrature constant
$A_{l},l=1,2,\cdots $as
\begin{equation}
\mu=\sum_{l} A_{l}\delta(x-x_{l}).
\end{equation}
Also, by using the recursion relation (\ref{6}) and quantum decomposition of adjacency matrix $A$ the other matrix element
$\langle\phi_k|A^m|\phi_0\rangle$ can be obtained as \cite{salimi}
\begin{equation}\label{cw1}
\braket{\phi_{k}}{A^m\mid
\phi_0}=\frac{1}{\sqrt{\omega_1\omega_2\cdots \omega_{k}
}}\int_{R}x^{m}P_{k}(x)\mu(dx),  \;\;\;\;\ m=0,1,2,....
\end{equation}
Finally, by using the above approach, we want to transfer quantum
states between antipodes of various networks perfectly. After
stratificating of each graph in respect to reference vertex $o$ and replacing $A$ whit $e^{-iH_{G}t}$,
we have
\begin{equation}\label{cw1}
\braket{\phi_{k}}{e^{-iH_{G}t}\mid
\phi_0}=\frac{1}{\sqrt{\omega_1\omega_2\cdots \omega_{k}
}}\int_{R}e^{-ixt}P_{k}(x)\mu(dx),  \;\;\;\;\ m=0,1,2,....
\end{equation}
where this relation is $f_{k,1}(t)$ and we use this point that adjacent matrix and Hamiltonian are
equivalent. Therefore, if we consider $k$ the antipode strata we can obtain the condition of perfect state transfer on spin networks.

\section {Examples}
In this section, we want to investigate perfect state transfer on
the variety of graphs by using the approaches given.

\subsection{Graph G$_n$}
 As the first example, the graphs taken from column method are
investigated. In column method, all vertices can be placed in $N$
column $G_{n}$ of size
 \begin{equation}
 |G_{n}|=\left(
\begin{array}{ccccc}
 N-1  \\
     n-1 \\
\end{array}
\right)
,\end{equation} that satisfy the following two conditions for n=1,2,\ldots,N:\\ (i) each vertex in column $n$ is connected to $N-n$ vertices in column n+1 \\(ii)each vertex in column n+1 is connected to $n$ vertices in column $ n$.\\
That Hamiltonian XY is defined as following
($J_{n,n+1}=J=1$)\cite{cdel},
\begin{equation}
H_{G}=\sum_{(i,j)\in E(G)}J_{i,j}(\sigma_{i}^{x}\sigma_{j}^{x}+
\sigma_{i}^{y}\sigma_{j}^{y})=\frac{1}{2}\sum_{(i,j)\in E(G)}
J_{i,j}(\sigma_{i}^{+}\sigma_{j}^{-}
+\sigma_{i}^{-}\sigma_{j}^{+}),
\end{equation}
 If $N=2$, i.e, only two vertices, for this formalism we have

$$|\phi_{0}\rangle=|1\rangle, \;\;\;\;\ |\phi_{1}\rangle=|2\rangle,$$
\begin{equation}
H=\left(
\begin{array}{ccccc}
 0 &  \frac{1}{2}  \\
      \frac{1}{2} & 0 \\

\end{array}
\right),
\end{equation} $$ \omega_{1}=\frac{1}{4}, \;\;\;\;\
\alpha_{1}=\alpha_{2}=\cdots \alpha_{n}=0,$$ \begin{equation}
P_{0}(x)=1, \;\;\;\;\ P_{1}(x)=x,\;\;\;\;\
P_{2}(x)=x^{2}-\frac{1}{4},
\end{equation}
And also

\begin{equation}
G_{\mu}(x)=\frac{x}{x^{2}-\frac{1}{4}},\;\;\;\;\
\ \mu(x)= \frac{1}{2} \delta(x-\frac{1}{2}) +\frac{1}{2}
\delta(x+\frac{1}{2}) ,
\end{equation}
Based on equation (\ref{cw1}),
 \begin{equation}
 f_{2,1}(t)=\langle \phi_{1}|e^{-i\lambda H_{G}t
}|\phi_{0}\rangle=\frac{1}{\sqrt{\omega_{1}}}\int e^{-i \lambda
xt}P_{1}(x) \mu (x) dx = -i\sin (\frac{\lambda t}{2} ) .
\end{equation}
If $N=4$ then ,we will have tree columns that one vertex in first
and third columns and two vertex in second   column, that degree of
each vertex is two. In result basic vectors  can be defined as
following:
$$|\phi_{0}\rangle=|1\rangle ,\;\;\;\;\ |\phi_{1}\rangle=\frac{1}{\sqrt{2}}(|2\rangle+|3\rangle) ,\;\;\;\;\ |\phi_{2}\rangle=|4\rangle,$$
 \begin{equation}H= \left(\begin{array}{ccccc}
 0 & \frac{1}{\sqrt{2}} & 0 \\
      \frac{1}{\sqrt{2}} & 0 & \frac{1}{\sqrt{2}}  \\
      0 &\frac{1}{\sqrt{2}} & 0 \\

\end{array}
\right),
\end{equation}$$
\omega_{1}=\frac{1}{2},\;\;\;\;\ \omega _{2}=
\frac{1}{2},\;\;\;\;\ \alpha _{1}=\alpha _{2}=\cdots \alpha_{n}=0
, $$  \begin{equation}   P_{0}(x)=1, \;\;\;\;\
P_{1}(x)=x,\;\;\;\;\ P_{2}(x)=x^{2}-\frac{1}{2},\;\;\;\;\
P_{3}(x)=x^{3}-x ,
\end{equation}
And also

 \begin{equation} G_{\mu}(x)
=\frac{x^{2}-\frac{1}{2}}{x^{3}-x },\;\;\;\;\ \mu (x) =\frac{1}{2}
\delta (x) +\frac{1}{4 } \delta (x-1) +\frac{1}{4} \delta (x+1),
\end{equation}
 Using equation (\ref{cw1}),
$$
 f_{3,1}(t)= \langle\phi_{2}|e^{-i\lambda
H_{G}t}|\phi_{0}\rangle=\frac{1}{\sqrt{\omega_{1}\omega_{2}}}\int
e^{-i\lambda  xt} P_{2}(x)\mu(x)dx,
$$
\begin{equation}
=\sin^{2}(\frac{\lambda t}{2})=-(-i\sin(\frac{\lambda t}{2}))^{2}
.\end{equation}
 If we consider $N=8$ then, we will have four column that one
vertex in first and forth column and three vertices in second and
third columns, that degree of each vertex is three. In result basic
vectors can be defined as following
$$|\phi_{0}\rangle=|1>,\;\;\;\;
|\phi_{1}>=\frac{1}{\sqrt{3}}(|2\rangle+|3\rangle+|4\rangle),\;\;\;\;|\phi_{2}\rangle=\frac{1}{\sqrt{3}}(|5\rangle+|6\rangle+|7\rangle),
\;\;\;\; |\phi_{3}\rangle=|8\rangle,$$
\begin{equation}
H= \left(\begin{array}{ccccc}
 0 & \frac{\sqrt{3}}{2} & 0& 0 \\
      \frac{\sqrt{3}}{2} &0 & 1 & 0  \\
      0 & 1& 0 & \frac{\sqrt{3}}{2}\\ 0& 0& \frac{\sqrt{3}}{2} &0\\
\end{array}
\right),
\end{equation}
\\ $$ \omega _{1}=\frac{3}{4}
,\;\;\;\;\ \omega_{2}=1 ,\;\;\;\;\ \omega_{3}
=\frac{3}{4},\;\;\;\;\ \alpha_{1}=\alpha_{2}=\cdots =\alpha_{n}=0
,$$
\begin{equation}
 P_{0}(x)=1,\;\;\;\;\ P_{1}(x)=x,\;\;\;\;\
P_{2}(x)=x^{2}-\frac{3}{4},\;\;\;\;\
P_{3}(x)=x^{3}-\frac{7}{4}x,\;\;\;\;\
P_{4}(x)=x^{4}-\frac{5}{2}x^{2}+\frac{9}{16},
\end{equation}
 so
\begin{equation}
G_{\mu}(x)=\frac{x^{3}-\frac{7}{4}x}{x^{4}-\frac{5}{2}x^{2}+\frac{9}{16}},\;\;\;\;\
\mu (x)= \frac{3}{8}\delta (x-\frac{1}{2})+\frac{1}{8}\delta
(x-\frac{3}{2})+ \frac{1}{8}\delta (x+\frac{3}{2})
+\frac{3}{8}\delta (x+\frac{1}{2}),
\end{equation}
Using equation (\ref{cw1}),
 \begin{equation}
 f_{4,1}(t)=\langle\phi_{3}|e^{-i\lambda
H_{G}t}|\phi_{0}\rangle=\frac{1}{\sqrt{\omega_{1}\omega_{2}\omega_{3}}}\int
e^{-i\lambda xt}P_{3}(x)\mu(x) dx =i\sin^{3}(\frac{\lambda
t}{2})=(-i \sin(\frac{\lambda t}{2}))^{3}.
\end{equation}
If $N=16$ then, we will have five columns that one vertex in first
and fifth columns, four vertex in second and fourth columns and sex
vertex in third columns. Degree of each vertex is four, in result
basic vectors can be defined as following:
$$|\phi_{0}\rangle=|1\rangle,$$ $$|\phi_{1}\rangle=\frac{1}{\sqrt{4}}(|2\rangle+|3\rangle+|4\rangle+|5\rangle),$$
$$ |\phi_{2}\rangle=\frac{1}{\sqrt{6}}(|6\rangle+|7\rangle+|8\rangle+|9\rangle+|10\rangle+|11\rangle),$$
$$|\phi_{3}\rangle=\frac{1}{\sqrt{4}}(|12\rangle+|13\rangle+|14\rangle+|15\rangle),$$  $$ |\phi_{4}\rangle=|16\rangle,  $$
\begin{equation}H= \left(\begin{array}{ccccc}
 0 & 1 & 0& 0&0 \\
     1& 0 &  \sqrt{\frac{3}{2}}&0 &0\\
      0 & \sqrt{\frac{3}{2}}& 0 &  \sqrt{\frac{3}{2}} &0 \\ 0& 0&\sqrt{\frac{3}{2}}&0&1 \\0& 0& 0&1 &0 \\

\end{array}
\right),
\end{equation}

$$  \omega_{1}=1 , \;\;\;\;\
\omega_{2}=\frac{3}{2},\;\;\;\; \omega_{3}=\frac{3}{2}\;\;\;\;,
\omega_{4 }=1,\;\;\;\;\alpha _{1}=\alpha_{2}=\cdots
\alpha_{n}=0,$$
\begin{equation}P_{0}(x)=1\;\;\;\;P_{1}(x)=x,\;\;\;\;P_{2}(x)=x^{2}-1,\;\;\;\;P_{3}(x)=x^{3}-\frac{5}{2}x,$$ $$\;\;\;\;\
P_{4}(x)=x^{4}-4x^{2}+\frac{3}{2},\;\;\;\;P_{5}(x)=x^{5}-5x^{3}+4x,\end{equation}
And also,
\begin{equation}
G_{\mu}(x)=\frac{x^{2}-4x+\frac{3}{2}}{x^{5}-5x+4x},\;\;\;\;\ \mu
(x)= \frac{1}{16}\delta (x-2)+\frac{1}{4}\delta
(x-1)+\frac{3}{8}\delta (x)+\frac{1}{4}\delta
(x+1)+\frac{1}{16}\delta (x+2),\end{equation} Based on equation
(\ref{cw1}),
\begin{equation}
f_{5,1}(t)=\langle\phi_{4}|e^{-i\lambda
H_{G}t}|\phi_{0}\rangle=\frac{1}{\sqrt{\omega_{1}\omega_{2}\omega_{3}\omega_{4}}}
\int e^{-i\lambda xt} P_{4}(x) \mu (x)=\sin^{4}(\frac{\lambda
t}{2})=(-i \sin(\frac{\lambda t}{2}))^{4}.
\end{equation}
If $t=\pi/\lambda$, then in all cases above, we have
$|f_{2,1}(t)|=|f_{3,1}(t)|=|f_{4,1}(t)|=|f_{5,1}(t)|=1$ that the
prefect quantum state transfer is obtained.

\subsection{W network}  Now, we want to investigate PST in case of
$W$ network. $W $ network is a eight-vertex
 and $E(W)={ \{ \{1,i\} , \{j,8\}2\leq i ,j\leq7 ; \}}$. We can
  consider this graph as a three-columns graph that there is one vertex in first and third column and are six vertex in second column.
 Vertex degree in second column is two and vertex degree in first and third is six. In this
 network, we have

$$ |\phi_{0}\rangle=|1\rangle, \;\;\ |\phi_{1}\rangle=\frac{1}{\sqrt{6}}(|2\rangle + |3\rangle + |4\rangle + |5\rangle + |6\rangle +
|7\rangle), \;\;\ |\phi_{2}\rangle=|8\rangle ,$$\
\begin{equation}
H= \left(\begin{array}{ccccc}
 0 & \sqrt{\frac{3}{2}} & 0 \\
 \sqrt{\frac{3}{2}} &0& \sqrt{\frac{3}{2}} \\0 & \sqrt{\frac{3}{2}}& 0  \\
\end{array}
\right),
\end{equation}
$$\omega_{1}=\frac{3}{2}, \;\;\;\; \omega_{1}=\frac{3}{2}, \;\;\;\; \alpha _{1}=\alpha_{2}=\cdots
\alpha_{n}=0,$$  \begin{equation}P_{0}(x)=1, \;\;\;\ P_{1}(x)=x,
\;\;\;\, P_{2}(x)=x^{2}-\frac{3}{2}, \;\;\;\;  P_{3}(x)=
x^{3}-3x,\end{equation} \\ so
 \begin{equation}G_{\mu}(x)=\frac{x^{2}-\frac{3}{2}}{x^{3}-3},\;\;\;\; \mu(x)=\frac{1}{2}\delta(x)
 +\frac{1}{4}\delta(x-\sqrt{3})
  +\frac{1}{4}\delta(x+\sqrt{3}),
 \end{equation}
Based on equation (\ref{cw1}),
\begin{equation}
f_{3,1}(t)=\langle\phi_{2}|e^{-i\lambda
H(G)t}|\phi_{0}\rangle=\frac{1}{\sqrt{\omega_{1}\omega_{2}}} \int
e^{-i \lambda xt} P_{2}(x)\mu (x)=-\sin^{2}(\frac{\sqrt{3} \lambda
t}{2}).
\end{equation}
 If we consider $t=\frac{\pi}{\sqrt{3}\lambda}$, we have prefect quantum state
  transfer $|f_{3,1}(t=\frac{\pi}{\sqrt{3}\lambda})|=1.$
 \subsection{Binary tree  network}
 Here, we want to investigate quantum state transfer in Binary
 tree network with $N=7$ vertices. In this network, we will have
 tree column that one vertex is in first column with two degrees
 and two vertices are in second column with tree degrees and four
 vertices in third column with one degree. Then we have
$$|\phi_{0}\rangle=|1\rangle, \;\;\;\;
|\phi_{1}\rangle=\frac{1}{\sqrt{2}}(|2\rangle+|3\rangle),
\;\;\;\;|\phi_{2}\rangle=\frac{1}{\sqrt{4}}(|4\rangle+|5\rangle+|6\rangle+|7\rangle),$$
\begin{equation}H= \left(\begin{array}{ccccc}
 0 & \frac{1}{\sqrt{2}} & 0 \\
      \frac{1}{\sqrt{2}} & 0 & \frac{1}{\sqrt{2}}  \\
      0 &\frac{1}{\sqrt{2}} & 0 \\

\end{array}
\right),
\end{equation}
$$\omega_{1}=\frac{1}{2},\;\;\;\;\ \omega _{2}=
\frac{1}{2},\;\;\;\;\ \alpha _{1}=\alpha _{2}=\cdots \alpha_{n}=0
, $$  \begin{equation}   P_{0}(x)=1, \;\;\;\;\
P_{1}(x)=x,\;\;\;\;\ P_{2}(x)=x^{2}-\frac{1}{2},\;\;\;\;\
P_{3}(x)=x^{3}-x ,
\end{equation}
also

 \begin{equation} G_{\mu}(x)
=\frac{x^{2}-\frac{1}{2}}{x^{3}-x },\;\;\;\;\ \mu (x) =\frac{1}{2}
\delta (x) +\frac{1}{4 } \delta (x-1) +\frac{1}{4} \delta (x+1).
\end{equation}
Using equation (\ref{cw1}),

\begin{equation}
f_{3,1}(t)=\langle\phi_{2}| e^{-i\lambda  H_{G} t}|\phi_{0}\rangle
=-\sin^{2}(\frac{ \lambda t}{2}).
\end{equation}
To obtain prefect quantum state transfer we consider
$t=\pi/\lambda$, i.e, $|f_{3,1}(t=\pi/\lambda)|=1$.

\subsection{Linear spin chain} In this example, we want to follow the
above discussion(PST) through a spin chain:
   $$H_{G}=\sum_{n=1}^{N-1}
J_{n,n+1}(\sigma_{n}^{x}\sigma_{n+1}^{x}+
\sigma_{n}^{y}\sigma_{n+1}^{y})=\frac{1}{2}\sum_{n=1}^{N-1}J_{n,n+1}(\sigma_{n}^{+}
\sigma_{n+1}^{-}\sigma_{n+1}^{+}),
$$
where  $J_{n,n+1}=\sqrt{n(N-n)}.$

If $N=2$, i.e.  total number of vertices is two, then $$
|\phi_{0}\rangle=|1\rangle, \;\;\;\ |\phi_{1}\rangle=|2\rangle,$$
\begin{equation}
H=\left(
\begin{array}{ccccc}
 0 &  \frac{1}{2}  \\
      \frac{1}{2} & 0 \\

\end{array}
\right),
\end{equation}

$$
\omega_{1}=\frac{1}{4}, \;\;\;\;\
\alpha_{1}=\alpha_{2}=\cdots \alpha_{n}=0,
$$
also
\begin{equation}
G_{\mu}(x)=\frac{x}{x^{2}-\frac{1}{4}},\;\;\;\;\ \ \mu(x)=
\frac{1}{2} \delta(x-\frac{1}{2}) +\frac{1}{2}
\delta(x+\frac{1}{2}),
\end{equation}
Based on equation (\ref{cw1}),
 \begin{equation}
 f_{2,1}(t)=\langle \phi_{1}|e^{-i\lambda H_{G}t
}|\phi_{0}\rangle=\frac{1}{\sqrt{\omega_{1}}}\int e^{-i\lambda
xt}P_{1}(x) \mu (x) dx= -i\sin (\frac{\lambda t}{2} ).
\end{equation}
If $N=3$ then
$$|\phi_{0}\rangle=|1\rangle ,\;\;\;\;\ |\phi_{1}\rangle=|2\rangle ,\;\;\;\;\
|\phi_{2}\rangle=|3\rangle,$$
 \begin{equation}H= \left(\begin{array}{ccccc}
 0 & \frac{1}{\sqrt{2}} & 0 \\
      \frac{1}{\sqrt{2}} & 0 & \frac{1}{\sqrt{2}}  \\
      0 &\frac{1}{\sqrt{2}} & 0 \\

\end{array}
\right),
\end{equation}$$
\omega_{1}=\frac{1}{2},\;\;\;\;\ \omega _{2}=
\frac{1}{2},\;\;\;\;\ \alpha _{1}=\alpha _{2}=\cdots \alpha_{n}=0
, $$
 \begin{equation}
  G_{\mu}(x)
=\frac{x^{2}-\frac{1}{2}}{x^{3}-x },\;\;\;\;\ \mu (x) =\frac{1}{2}
\delta (x) +\frac{1}{4 } \delta (x-1) +\frac{1}{4} \delta (x+1),
\end{equation}
\begin{equation}
f_{3,1}(t)=\langle\phi_{2}|e^{-i\lambda
H_{G}t}|\phi_{0}\rangle=\frac{1}{\sqrt{\omega_{1}\omega_{2}}}\int
e^{-i\lambda xt} P_{2}(x)\mu(x)dx =\sin^{2}(\frac{\lambda
t}{2})=-(-i\sin(\frac{\lambda t}{2}))^{2}
.\end{equation}\\

if $N=4$ then$$|\phi_{0}\rangle=|1\rangle,\;\;\;\;
|\phi_{1}\rangle=|2\rangle,\;\;\;\;|\phi_{2}\rangle=|3\rangle,
\;\;\;\; |\phi_{3}\rangle=|4\rangle,$$
\begin{equation}H= \left(\begin{array}{ccccc}
 0 & \frac{\sqrt{3}}{2} & 0& 0 \\
      \frac{\sqrt{3}}{2} &0 & 1 & 0  \\
      0 & 1& 0 & \frac{\sqrt{3}}{2}\\ 0& 0& \frac{\sqrt{3}}{2} &0\\

\end{array}
\right),
\end{equation} \\ $$ \omega _{1}=\frac{3}{4}
,\;\;\;\;\ \omega_{2}=1 ,\;\;\;\;\ \omega_{3}
=\frac{3}{4},\;\;\;\;\ \alpha_{1}=\alpha_{2}=\cdots =\alpha_{n}=0
,$$
\begin{equation}
G_{\mu}(x)=\frac{x^{3}-\frac{7}{4}x}{x^{4}-\frac{5}{2}x^{2}+\frac{9}{16}},\;\;\;\;\
\mu (x)= \frac{3}{8}\delta (x-\frac{1}{2})+\frac{1}{8}\delta
(x-\frac{3}{2})+ \frac{1}{8}\delta (x+\frac{3}{2})
+\frac{3}{8}\delta (x+\frac{1}{2}).
\end{equation}
Using equation (\ref{cw1})
\begin{equation}
f_{4,1}(t)=\langle\phi_{3}|e^{-i\lambda
H_{G}t}|\phi_{0}\rangle=\frac{1}{\sqrt{\omega_{1}\omega_{2}\omega_{3}}}\int
e^{-i\lambda xt}P_{3}(x)\mu(x) dx =i\sin^{3}(\frac{\lambda
t}{2})=(-i \sin(\frac{\lambda t}{2}))^{3}.
\end{equation}
 If
$N=5$,$$|\phi_{0}\rangle=|1\rangle,\;\;\;\;
|\phi_{1}\rangle=|2\rangle,\;\;\;\;\
 |\phi_{2}\rangle=|3\rangle,\;\;\;\;\
|\phi_{3}\rangle=|4\rangle,\;\;\;\;\ |\phi_{4}\rangle=|5\rangle,  $$
\begin{equation}H= \left(\begin{array}{ccccc}
 0 & 1 & 0& 0&0 \\
     1& 0 &  \sqrt{\frac{3}{2}}&0 &0\\
      0 & \sqrt{\frac{3}{2}}& 0 &  \sqrt{\frac{3}{2}} &0 \\ 0& 0&\sqrt{\frac{3}{2}}&0&1 \\0& 0& 0&1 &0 \\

\end{array}
\right),
\end{equation}

$$  \omega_{1}=1 , \;\;\;\;\
\omega_{2}=\frac{3}{2},\;\;\;\; \omega_{3}=\frac{3}{2}\;\;\;\;,
\omega_{4 }=1,\;\;\;\;\alpha _{1}=\alpha_{2}=\cdots \alpha_{n}=0,$$
also
\begin{equation}
G_{\mu}(x)=\frac{x^{2}-4x+\frac{3}{2}}{x^{5}-5x+4x},\;\;\;\;\ \mu
(x)= \frac{1}{16}\delta (x-2)+\frac{1}{4}\delta
(x-1)+\frac{3}{8}\delta (x)+\frac{1}{4}\delta
(x+1)+\frac{1}{16}\delta (x+2),
\end{equation}
Based on equation (\ref{cw1})
\begin{equation}
f_{5,1}(t)=\langle\phi_{4}|e^{-i\lambda
H_{G}t}|\phi_{0}\rangle=\frac{1}{\sqrt{\omega_{1}\omega_{2}\omega_{3}\omega_{4}}}
\int e^{-i\lambda xt} P_{4}(x) \mu (x)=\sin^{4}(\frac{\lambda
t}{2})=(-i \sin(\frac{\lambda t}{2}))^{4},
\end{equation}
and if we continue this process, we will have [4],
\begin{equation}
f_{N,1}(t)=\langle\phi_{N-1}|e^{-i\lambda H_{G}t}|\phi_{0}\rangle=
\left[-i \sin(\frac{\lambda t}{2} )\right]^{N-1}.
\end{equation}
In result, in  $t=\pi/\lambda$ we have perfect state transfer
between antipodal, i.e,
$$|f_{N,1}(t=\pi/\lambda)|=1.$$
In following cases, with change $J_{n,n+1} $ we can transfer  quantum  states
 between antipodal perfectly.\\
\subsection{Star network}
We consider star network with $N=5$. In this network, one vertex
with one degree is in first column and one vertex with five degree
is in second column and three vertices with one degree  are in third
column. Also in this case we have
\begin{equation}J_{1,2}=\sqrt{3},\;\;\;\,J_{2,3}=J_{2,4}=J_{2,5}=1,
\end{equation}
 $$|\phi_{0}\rangle=|1\rangle
,\;\;\;\;|\phi_{1}\rangle=|2\rangle,\;\;\;\;|\phi_{2}\rangle=\frac{1}{\sqrt{3}}(|3\rangle+|4\rangle+|5\rangle),$$
\begin{equation}
H=\left(
\begin{array}{ccccc}
 0 &  \frac{\sqrt{3}}{2} & 0 \\
    \frac{\sqrt{3}}{2}& 0 & \frac{\sqrt{3}}{2}   \\ 0 &
    \frac{\sqrt{3}}{2}&0 \\

\end{array}
\right),
\end{equation}

$$ \omega_{1}=\omega_{2}=\frac{3}{2},\;\;\;\;\alpha_{1}=\alpha_{2}=\cdots=\alpha_{n}=0$$
\begin{equation} P_{0}=1,\;\;\;\;P_{1}=x,\;\;\;\;P_{2}=x^{2}-\frac{3}{4},\;\;\;\;P_{3}=x^{3}-\frac{3}{2}x
,\end{equation}
and also
\begin{equation}
 G_{\mu}(x)=\frac{x^{2}-\frac{3}{4}}{x^{3}-\frac{3}{2}x},\;\;\; \;\;\;\;
 \\\mu(x)=\frac{1}{2}\delta(x) +\frac{1}{4}\delta(x-\sqrt{\frac{3}{2}})+\frac{1}{4}\delta(x+\sqrt{\frac{3}{2}}),
 \end{equation}
\begin{equation}
f_{3,1}(t)=\langle\phi_{2}|e^{-i\lambda
H_{G}t}|\phi_{0}\rangle=\frac{1}{\sqrt{\omega_{1}\omega_{2}}}\int
e^{-i\lambda xt}P_{2}(x)\mu
(x)dx=-\sin^{2}(\sqrt{\frac{3}{2}}\lambda t).
\end{equation}
In result, for $ t= \sqrt{\frac{2}{3}}\pi/\lambda$ we have perfect
state
transfer between antipodal.\\
\subsection{Circulant network}
Circulant network with $N=6$ is define as following: one vertex is
in first and fourth column and two vertices are in second and third
column that degree of each vertex in this network is two. For this
network, we have
\begin{equation}J_{1,2}=J_{1,3}=\sqrt{\frac{3}{2}},\;\;\;\;J_{2,4}=J_{3,5}=2, \;\;\;\;J_{4,6}=J_{5,6}=\sqrt{\frac{3}{2}},\end{equation}
$$|\phi_{0}\rangle=|1\rangle, \;\;\;\;|\phi_{1}\rangle=\frac{1}{\sqrt{2}}(|2\rangle+|3\rangle),\;\;\;\;
|\phi_{2}\rangle=\frac{1}{\sqrt{2}}(|4\rangle+|5\rangle)\;\;\;\;|\phi_{3}\rangle=|6\rangle,$$
\begin{equation}H= \left(\begin{array}{ccccc}
 0 & \frac{\sqrt{3}}{2} & 0& 0 \\
      \frac{\sqrt{3}}{2} &0 & 1 & 0  \\
      0 & 1& 0 & \frac{\sqrt{3}}{2}\\ 0& 0& \frac{\sqrt{3}}{2} &0\\
\end{array}
\right),
\end{equation} \\
$$ \omega _{1}=\frac{3}{4}
,\;\;\;\;\ \omega_{2}=1 ,\;\;\;\;\ \omega_{3}
=\frac{3}{4},\;\;\;\;\ \alpha_{1}=\alpha_{2}=\cdots =\alpha_{n}=0
,$$
 \begin{equation}
  P_{0}(x)=1,\;\;\;\;\ P_{1}(x)=x,\;\;\;\;\
P_{2}(x)=x^{2}-\frac{3}{4},\;\;\;\;\
P_{3}(x)=x^{3}-\frac{7}{4}x,\;\;\;\;\
P_{4}(x)=x^{4}-\frac{5}{2}x^{2}+\frac{9}{16},
\end{equation}
so
\begin{equation}
G_{\mu}(x)=\frac{x^{3}-\frac{7}{4}x}{x^{4}-\frac{5}{2}x^{2}+\frac{9}{16}},\;\;\;\;\
\mu (x)= \frac{3}{8}\delta (x-\frac{1}{2})+\frac{1}{8}\delta
(x-\frac{3}{2})+ \frac{1}{8}\delta (x+\frac{3}{2})
+\frac{3}{8}\delta (x+\frac{1}{2}),
\end{equation}
Using equation (\ref{cw1})
 \begin{equation}
 f{4,1}(t)=\langle\phi_{3}|e^{-i\lambda
H_{G}t}|\phi_{0}\rangle=\frac{1}{\sqrt{\omega_{1}\omega_{2}\omega_{3}}}\int
e^{-i\lambda xt}P_{3}(x)\mu(x) dx=i\sin^{3}(\frac{\lambda
t}{2}).
\end{equation}
 In result, for $t=\pi/\lambda$ we have perfect state
transfer.
\subsection{Binary tree network with modulating coupling}
In this example, we consider binary tree spin networks with $N=16$
vertices with
$$
J_{1,2}=2,\;\;\;\;\ J_{2,3}=J_{2,4}=\sqrt{3},\;\;\;\;\ J_{3,5}=J_{3,6}=J_{4,7}=J_{4,8}=\sqrt{3}
$$
\begin{equation}
J_{5,9}=J_{5,10}=J_{6,11}=J_{6,12}=J_{7,13}=J_{7,14}=J_{8,15}=J_{8,16}=\sqrt{2}.
\end{equation}
For this graph. we have
 $$ |\phi_{0}\rangle=|1\rangle,\;\;\;\;\
|\phi_{1}\rangle=|2\rangle,\;\;\;\;\
|\phi_{2}\rangle=\frac{1}{\sqrt{2}}(|3\rangle+|4\rangle),$$
$$|\phi_{3}\rangle=\frac{1}{\sqrt{4}}(
|5\rangle+|6\rangle+|7\rangle+|8\rangle) +$$
$$\frac{1}{\sqrt{8}}(|9\rangle+|10\rangle+|11\rangle+|12\rangle+|13\rangle+|14\rangle+|15\rangle+|16\rangle),$$
where the form of Hamiltonian in the above basis is
\begin{equation}
H= \left(\begin{array}{ccccc}
 0 & 1 & 0& 0&0 \\
     1& 0 &  \sqrt{\frac{3}{2}}&0 &0\\
      0 & \sqrt{\frac{3}{2}}& 0 &  \sqrt{\frac{3}{2}} &0 \\ 0& 0&\sqrt{\frac{3}{2}}&0&1 \\0& 0& 0&1 &0 \\

\end{array}
\right),
\end{equation}
$$  \omega_{1}=1 , \;\;\;\;\
\omega_{2}=\frac{3}{2},\;\;\;\; \omega_{3}=\frac{3}{2}\;\;\;\;,
\omega_{4 }=1,\;\;\;\;\alpha _{1}=\alpha_{2}=\cdots
\alpha_{n}=0,$$
\begin{equation}P_{0}(x)=1\;\;\;\;P_{1}(x)=x,\;\;\;\;P_{2}(x)=x^{2}-1,\;\;\;\;P_{3}(x)=x^{3}-\frac{5}{2}x,$$ $$\;\;\;\;\
P_{4}(x)=x^{4}-4x^{2}+\frac{3}{2},\;\;\;\;P_{5}(x)=x^{5}-5x^{3}+4x,
\end{equation}
 also
\begin{equation}
G_{\mu}(x)=\frac{x^{2}-4x+\frac{3}{2}}{x^{5}-5x+4x},\;\;\;\;\ \mu
(x)= \frac{1}{16}\delta (x-2)+\frac{1}{4}\delta
(x-1)+\frac{3}{8}\delta (x)+\frac{1}{4}\delta
(x+1)+\frac{1}{16}\delta (x+2).
\end{equation}
Based on equation (\ref{cw1})
\begin{equation}
f_{5,1}(t)=\langle\phi_{4}|e^{-i\lambda
H_{G}t}|\phi_{0}\rangle=\frac{1}{\sqrt{\omega_{1}\omega_{2}\omega_{3}\omega_{4}}}
\int e^{-i\lambda xt} P_{4}(x) \mu (x)=\sin^{4}(\frac{\lambda t}{2}).
\end{equation}
In result, for $t=\pi/\lambda$ we have perfect state transfer
between antipodal.
\section{Conclusion}
In this paper, we have focused on perfect quantum states transfer between antipodes of
 networks i.e, fidelity equivalent one. Thus, we have presented a total approach for perfect
quantum states transfer and by using this, we were able to transfer
quantum states in column networks, W network, spin chain, star
network,$\cdots$ perfectly. By using this approach, we can transfer
quantum    states in various network between antipodes with correct
choice $J_{m,n}$ perfectly.  Advantage of    this approach have is
that there is no need to have complicated eigenvalue and eigenvector
computations. For example in column network with $16$ vertices, we
have a $5\times5$ Hamiltonian instead of a $16\times16$ Hamiltonian.
\section{Acknowledgements}
We would like to thank A. R. J. Azar and D. Karami for help comments on an earlier manuscript.

\end{document}